\documentclass[12pt]{article}

\usepackage{graphicx}

\begin{document}

\title{Quantum Degrees of Freedom, Quantum Integrability and Entanglment Generators}
 \author{
 Nikola Buri\' c\thanks{e-mail: buric@phy.bg.ac.yu}
 \\  Institute of Physics, University of Belgrade,\\ PO BOX 68, 11000 Belgrade,
 Serbia.}

 \maketitle

\begin{abstract}

Dynamical algebra notion of quantum degrees of freedom is utilized
to study the relation between quantum dynamical integrability
 and generalized entanglement. It is argued that a quantum dynamical
 system generates generalized entanglement by internal dynamics if
 and only if it is quantum non-integrable. Several examples are
 used to illustrate the relation.

\end{abstract}


\newpage

Quantum theory is generally considered as the fundamental theory
of Physics. This means, among other things, that all physical
notions
 that appear in it should be derived within the quantum theory itself, that is without reference to another independent physical theory.
In this note we shall analyze the notion of an independent degree
of freedom (IDF) as it might be (and has been) defined
 in quantum mechanics, and study two important issues, seemingly unrelated, where the definition of quantum mechanical degrees of freedom is relevant.
i.e. the definition of entanglement and of quantum integrability.

IDF are relevant to the question of how much a state of a given
system possess the properties that are typically quantum. Total
level of quantum fluctuations  with respect to all basic
observables, which depend on the IDF of the quantum system, should
be considered as a measure of
 quantumness of a given quantum state.
 It is well known that a pure state of a fixed quantum system considered in differently defined sets of IDF
 might show different properties of entanglement.
 Maximally entangled state in one set of coordinates corresponding to one set of degrees of freedom
 might be disentangled and posses minimal quantumness in another set of degrees of freedom.

Another property of a quantum system which depends on the definition of IDF is the notion of quantum integrability.
 A unique notion of the quantum dynamical integrability is not commonly accepted as it is in the classical case, and, like entanglement,
 depends on the definition of the IDF. A choice of IDF is possible such that any quantum system with a finite dimensional Hilbert space can
 be considered as a completely
 integrable Hamiltonian dynamical system.
 However, similar in the spirit and the meaning of the
 classical integrability is the notion of dynamical symmetry based on the systems dynamical algebra,
  which is well defined in quantum as well as in the classical mechanics.
 The notion of dynamical symmetry, and more generally the dynamical group, in quantum mechanics implies a well defined notion of
 independent dynamical degrees of freedom, and can be used to define quantum integrability. Although quantum systems never display
 qualitative properties of classical nonintegrable systems
 a  quantum system
 which is quantum non-integrable according to this definition has well defined classical model which shows typical qualitative properties
 of chaotic classical Hamiltonian systems.
On the other hand, the classical model of a quantum integrable
system is completely integrable in the classical sense.

 The IDF are determined by the need to describe the interactions within the system and possible interactions of the considered system and
 the environment which serve to describe what can be measured in the given circumstances.
  For example, if the system is composed of two spatially separated subsystems than it is "natural for local observers" to choose
 the independent degrees of freedom to respect the spacial separation.
 If the spacial separation leads also to dynamical independence of the two subsystems then
 an arbitrary separable state of such a system remains separable in the course of
 the evolution, and this is an objective property of the system.
 In this case the two sets of entangled and separable states are dynamically separated. Thus, it could be argued that the
 choice of
 IDF in this case appears natural precisely because of the dynamical separation between entangled and separable states. This suggest
 that in general the quantum definition of IDF that displays the objective property of entanglement
 is related to the quantum dynamical properties of the system i.e. to the quantum integrability.

Our goal will be to study the importance of the definition of IDF for
 the relation between entanglement and quantum integrability for a quantum dynamical system in general.
We shall see that in quantum mechanics the choice of degrees of
freedom dictated by the dynamical structure of the system that is
by the dynamical algebra and its particular subalgebras should
represent also an appropriate choice
 of IDF for an objective and generalized treatment of entanglement. Discussion of the systems dynamical group will lead to a
 notion of degrees of freedom and the discussion of dynamical symmetry to the notion of quantum non-integrability and entanglement generating systems.
 In a nutshell, our conclusion will be that
 the dynamical group of the system determines its degrees of freedom and the dynamical symmetry determines if the
 quantum system is capable of generating entanglement.

In the next section we shall first recapitulate general
definitions of independent degrees of freedom, quantum
integrability and the generalized entanglement. We then establish
the relation between the quantum integrability and generalized
entanglement and discuss, in section 3, several examples.
 Dynamical
algebraic definition of IDF and quantum integrability have been
introduced in references \cite{Cheng1,Cheng2,Cheng3}.
 There is no generally accepted notion of genuinely quantum
  integrability \cite{Bethe}. The definition of what is a quantum chaotic system is
  even less unique \cite{Semi-group}. The most common approach, at least for lattice spin
  systems, is based on the generalization of the notion of
  thermodynamical integrability of classical spin systems\cite{Baxter},
   and is different from the one accepted here.
   According to the thermodynamic integrability a quantum system is called
   integrable if it is exactly solvable by application of the
   generalized Bethe ansatz or by the quantum inverse scattering
   method \cite{Bethe}.
    A quantum system is nonintegrable if it has not been
   integrated by such methods. In what sense a quantum nonintegrable
   system can be considered as quantum chaotic is a matter of a debate \cite{Prozen}.
   Some quantum systems of finite number of spins whose
   thermodynamical limit is quantum nonintegrable, show the same
   spectral properties as the systems obtained by quantization of
   classically chaotic systems, and, furthermore, display the mixing properties
    that lead to expected equilibrium and non-equilibrium thermodynamical behavior \cite{Prozen}.
   The dynamics of bipartite and multipartite entanglement in
   such quantum chaotic systems has been studied and compared with
   the entanglement dynamics in quantum integrable systems
   \cite{indusi1},\cite{indusi2},\cite{Casati},\cite{JaPhysLett}
The notion of quantum integrability and nonintegrability
understood in the thermodynamic sense is very different from the
notion of  dynamical symmetry and quantum integrability as was
introduced in \cite{Cheng1,Cheng2,Cheng3} and as it shall be used
here.

 The definition of generalized entanglement adopted here was presented
in \cite{Viola1,Viola2,Viola3}, and related definitions appear for
example in \cite{Klyasko},\cite{Zanardi}.   Here presented view of
the general relation between quantum integrability, dynamics of
classical approximations of quantum systems and dynamics of the
generalized entanglement has not been discussed before.

\section{General definitions}

\subsection{ Dynamical algebra framework}

{\it Kinematical degrees of freedom}:

 Any quantum system with an
N-dimensional Hilbert space has $N-1$ kinematical degrees of
freedom.
 Its group of canonical transformations,  i.e. the kinematical group, is $U(N)$ so that any Hamiltonian, i.e. Hermitian operator,
 can be digitalized using some of the $U(N)$ transformations,
 leading to $N-1$ formal integrals of motion and formal integrability. Evolution of the quantum system is equivalent to a linear symplectic flow
 on a symplectic manifold, which is completely integrable in the sense of classical Hamiltonian systems (please see \cite{AnnPhys} and the
  references therein).
 All pure states can be connected by some unitary transformation so that all
 pure states are in fact $U(N)$ generalized coherent states.
 The assumption that any hermitian operator represents a measurable quantity,  an observable, is actually an assumption concerning
 physically possible interactions with the environment, and is not justified in many cases such as: systems of identical particles, presence of
 symmetries, relativistic locality etc....

{\it Dynamical algebra of relevant observables}

 The notion of K-freedom is to formal to be physically relevant. A particular physical system is specified, and thus distinguished from
 an abstract general framework, by describing what can be measured on it, i.e. by specifying the set of observables,
 and by expressing the interactions within the system in terms of the observables.
In order words, the class of physically relevant observables
should be described and the evolution should be expressed in terms
of these observables. Structure of the set of observables is fixed
by their algebraic relations. In quantum mechanics operators
representing the physical quantities pertinent to the given system
 are required to realize the corresponding algebra. The algebraic relations between the observables also fix the relevant Hilbert space
 of the system as the space of an irreducible representation. The algebra defined in this way is called the systems dynamical algebra.
 Thus, a quantum system has fixed dynamical algebra.
 Description of a quantum system amounts to the specification of its state space, algebra of observables, or the dynamical algebra g,
 and the Hamiltonian which is an expression (possibly nonlinear) in terms of operators belonging to g.

In what follows we shall consider a quantum system $(H,g,{\cal
H})$ with a Hilbert space $H$ which is  an irrep space of the
dynamical (Lie) algebra $g$
 and the Hamiltonian ${\cal H}$.
The dynamical algebra $g$ will always be a semi-simple Lie
algebra, with rank $l$ and dimension $n$.

\subsection{ Dynamical degrees of freedom}

Dynamical degrees of freedom are fixed by the full description of
the system, and are defined using the dynamical algebra. The
dynamical algebra $g$ has $\gamma$ different chains of
subalgebras: $g\supset g^l_{s^l}\supset g^{l}_{s^l-1}\dots \supset
g^{l}_{1},\>l=1,2,\dots \gamma$.
 Casimir operators
 of $g$ and all the algebras in (any of) the subalgebra chain form the relevant complete set of commuting operators (CSCO) $Q_j, j=1,2\dots d$.
 There is $d=l+(n-l)/2$ of these,
 independently of the subalgebra chain.
Some of these Casimir operators are fully degenerate in the sense
that they are represented by scalar operators:
 $Q_i|\psi>=c_i|\psi>$ for every $|\psi>\in H$.
The number of non-fully degenerate operators in CSCO is $m\leq
(n-l)/2$ is chain independent but might depend on the particular
irrep i.e. on the system's Hilbert space, and defines the number
of IDF. The non-fully degenerate operators in a particular chain
are the operators that define  $m$ IDF.
 The quantum system is fully specified only when $1^o$ its Hilbert space; $2^o$ the set of $m$ operators representing IDF and
 $3^o$ the Hamiltonian, which is a possibly nonlinear expression in terms of the dynamical algebra generators, are given.

If the dynamical algebra $g$ of a quantum system $C$ can be represented as a direct sum of dynamical algebras of two systems $A$ and $B$,
 that is $g^C= g^A\oplus g^B$, then the tensor product of irreps of $G^A$ and $G^B$ is an irrep space of $G^C$, that is $H^C=H^A\otimes H^B$.
If $l_{A,B}$ and $n_{A,B}$ are the ranks and dimensions of  $g^A$
and $g^B$, then in general the number of IDF of $C$ is
$M_C=M_A+M_B$. Thus, in the case $g^C= g^A\oplus g^B$ the system
$C$ can be represented as a union of two  systems and the number
of IDF is additive. If the dynamical algebra $g$ of the system is
semisimple then it can be uniquely expressed as a direct sum of
mutually commutative
 and orthogonal simple algebras: $g=\oplus_k g_k$ and
the Hilbert space which is an irrep space of $g$ factors as
$H=\otimes_k H_k$. Thus, in the case of semisimple dynamical
algebra the number of IDF
 is additive, but the number of IDF in all the factor systems with $g_k$ dynamical algebras need not be unity for each $g_k$.
An example of the system when this is the case is given by a system of qubits, and shall be treated in some detail later.

However, a dynamical algebra $g$ need not be representable as a
product of dynamical Lie algebras with the number of IDF equal to
one,
 as for example if $g=su(3)$ or if
 $A$ and $B$ are independent fermions or bosons, even if the number of IDF of $g$
 is larger then one (for example in the $su(3)$ case it is 2 or 3 depending on the irrep, as discussed in section 3.4).

\subsection{ Dynamical symmetry i.e. integrability}

$(H,g, {\cal H})$ has the corresponding Lie group $G$ as the dynamical symmetry if the Hamiltonian  ${\cal H}$ can be expressed in terms of the CSCO
 of a particular subgroup chain used to define the IDF. In this case the system has a symmetry of the subgroup chain.
In particular $H$ commutes with $m$ non-fully degenerate operators
that define the IDF.

A system $(H,g, {\cal H})$  is quantum integrable by definition if it has $G$ dynamical symmetry with respect to the subgroup chain that is
 used to define the IDF.

$G$ symmetry is defined as quantum integrability in analogy with
 complete integrability in the presence of symmetry in the case of classical Hamiltonian systems.
Quantum Hamiltonian systems which do not satisfy the definition of
 quantum integrability are called quantum nonintegrable. It should be stressed that the qualitative properties
 of the state dynamics with quantum integrable and quantum non-integrable Hamiltonians are the same.
 From the point of view of the Hamiltonian dynamical systems theory the state orbits are in either case
regular that is  periodic or quasi-periodic. Quantum non-integrable systems do not generate chaotic orbits in
 the system's state space (please see for example \cite{AnnPhys}).
Nevertheless, the dynamical properties of orbits of the classical
models (please see the next subsection) corresponding to the
quantum integrable or non-integrable systems are quite
 different, and chaotic orbits do occur in the classical model of the quantum non-integrable systems with more than one IDF.

It should be noticed that quantum systems with one degree of freedom, unlike the one freedom classical Hamiltonian systems,
 need not be quantum integrable, for example if the Hamiltonian is a nonlinear expression of the algebra generators.

\subsection{ g-coherent states}

Total level of quantum fluctuations in a pure state $|\psi>$ is defined as
\begin{equation}
\Delta(\psi)=\sum_i^n<\psi|L_i^2|\psi>-<\psi|L_i|\psi>^2,
\end{equation}
where the sum is taken over an orthonormal bases of the dynamical
algebra $g$. It make sense to consider the quantity $\Delta(\psi)$
as a measure of  quantumness of the state $\psi$.
 Physical motivation for the definition of the generalized $g$-coherent states is that they minimize
 $\Delta(\psi)$. This is one of the important properties of the Glauber coherent states of the harmonic oscillator i.e.
 of the Haisenberg-Weil $H_4$ algebra that is generalized by the $g$-coherent states with arbitrary $g$.
 There are several  generalizations of the Glauber, i.e. $H_4$ coherent states.
 Perelomov \cite{Perelomov}
 and Gillmore \cite{Gilmor} independently introduced two different generalizations based on the group-theoretical structure of the $H_4$ coherent states.
  The essential
 ideas of both approaches are the same, the differences being in the class of Lie groups, and the corresponding available tools, and in
  the choice a reference state.
In both approaches, the set of $g$-coherent states depends,
besides the algebra $g$, also on the particular Hilbert space
$H^{\Lambda}$ caring the irrep $\Lambda$ of $g$ and on the choice
 of an, in principal (Perelomov), arbitrary referencee state, denoted $|\psi_0>$. The subgroup $S_{\psi_0}$ of $G$ which
  leaves the ray corresponding to the state $|\psi_0>$ invariant is called the stability subgroup of
 $|\psi_0>$: $h|\psi_0>=|\psi_0>exp i\chi(h), h\in S_{\psi_0}.$
 Then, for every $g\in G$ there is a unique
 decomposition into the product of two elements, one from $S_{\psi_0}$ and one from the coset $G/S_{\psi_0}$ so that
  $g|\psi_0>= \Omega|\psi_0> exp i\chi(h)$.
 The states of the form $|\Lambda, \Omega>=\Omega|\psi_0>$ for all $g\in G$ are the $g$ coherent states.
 Thus, geometrically the set of $g$ coherent state
 form a manifold with well defined Riemanien and symplectic structure.

 In all explicit examples treated here the dynamical algebra $g$ will be  semisimple (or simple), which is the case studied by Gillmore.
 In this case there is the standard Cartan basis of $g$: $\{H_i,E_{\alpha},E_{-\alpha}\}$. The irrep is characterized by the unique
 highest weight state $|\Lambda,\Lambda>$ (or the lowest weight state $|\Lambda,-\Lambda>$)
  which is annihilated by all $E_{\alpha}$ and some $E_{-\alpha}$.
  The state  $|\Lambda,\Lambda>$ is left invariant by  operators in the Cartan subalgebra $H_i$.    The set of $g$ coherent states
 can be represented in the form of an action of the so called displacement operator
 on the reference state $|\Lambda,\Lambda>$.
\begin{equation}
|\alpha>=D(\alpha)|\Lambda,\Lambda>=\exp[\sum
\alpha_iE_i-h.c.]|\Lambda,\Lambda>,
\end{equation}
where $\alpha_i$ are complex parameters and the sum extends over
all $E_{-\alpha}$ that do not annihilate $|\Lambda,\Lambda>$. The
stabilizer $S_{\psi_0}$ of the reference state $|0>$ is the
subgroup
 generated by the Cartan subalgebra of $g$
The complex parameters $\alpha_i,i=1,2\dots M$ parameterize $2M$
dimensional manifold $G/S_{\psi_0}$.

{\it Classical model and semi-classical dynamics}

Classical Hamiltonian dynamical system on the manifold
$G/S_{|\psi_0>}$ given by the Hamiltonian function ${\cal
H}(\alpha)=<\alpha|{\cal H}|\alpha>$ is called the classical model
of the quantum dynamical system $(H,g,{\cal H})$. Classical limit
of the quantum system is obtained from the classical model in the
limit when some relevant parameter approaches zero. If the
Hamiltonian ${\cal H}$ is a linear expression of the dynamical
group generators then the quantum system, its classical model and
its classical limit have the same dynamics. The classical model of
a quantum nonintegrable system is chaotic in the sense of
classical Hamiltonian dynamical systems. Dynamics of classical
models of quantum nonintegrable systems have been studied for
various examples in \cite{Cheng1,Cheng2,Cheng3}. Relation between
dynamics of entanglement and the dynamics of classical models for
quantum nonintegrable pair of qubits was studied in
\cite{JaPhysRev06}.

Dynamics of the traditional semi-classical approximation of the
quantum systems which are based on quantization of systems with
different classical dynamics has been studied intensively\cite
{Haake}.
   It is found that a) Quantum systems
  obtained by quantization of classical Hamiltonian system with
  qualitatively different dynamics show different spectral
  properties, and qualitatively different properties of entanglement in eigenstates
  in different parts of spectra have been observed \cite{Casati}; b)
  Bipartite standard entanglement, as measured by concurrence, in the wave function initially localized in
  qualitatively different parts of the phase space of some semi-classical
  approximation of the quantum system has clearly different
  dynamics ( for example \cite{Furuya, Miller, Shep1,Shep2, Zanardi2, Laksh,Tanaka,Laksh3, Hu}).

\subsection{ Generalized entanglement}

Consider a system $C$ such that its dynamical group $G^C$ can be factored as a direct product
$G^C=G^A\otimes G^B$, and its Hilbert space $H^C$ written as the tensor
 product of the irreps $H^C=H^A\otimes H^B$.
 We have seen that such a system $C$ can be viewed as a union of systems $A$ and $B$. A pure state $\psi_C$
 of $C$ is entangled by the standard definition if the reduced states $\rho_{A,B}=Tr_{B,A}[|\psi_C><\psi_C|]$ are mixtures i.e. are not  projectors.

 g-coherent states of the system $C$ with $G^C=G^A\times G^B$  are products of $g^A$ and $g^B$ coherent states, by the construction of coherent states
with the referent state $\psi^C_0=|\psi^A_0>\otimes|\psi^B_0>$ and
are of the form $|\alpha^A>\otimes |\alpha^B>=G^A|\psi^A_0>\otimes
G^B|\psi_0^B>$.
 Reduced states $\rho_{A,B}$ of the coherent state $|\alpha^A>\otimes |\alpha^B>$ are pure and are coherent states of $A$ and of $B$ respectively.
 The coherent states of $G^C$ are disentangled, and the reduced state of the coherent state are also coherent,
  and thus disentangled for the component algebras. Thus, the set of states with zero entanglement and the set of coherent states can
  be consistently identified. In this sense the
noncoherent states do posses some entanglement in the generalized
sense. If $A$ and $B$ are systems with only one IDF each, the
previous definition assumes
 that the noncoherent states of $A$ and $B$ are entangled in the generalized sense.
  These states $|\psi^A>,|\psi^B>$ of systems $A$ and $B$ with number of IDF equal to unity
   do have nonminimal quantumness $\Delta(\psi)$, and violate a Bell inequality
   for some set of observables \cite{Klyasko}.

In the considered case the quantumness $\Delta(\psi^C)$ is in
general larger then minimal, the minimum being achieved by states
which are products of $G^A$ and $G^B$ coherent states. The
quantumness of the state $|\psi^C>$ is here manifested in one of
the two modes: a) by quantum correlations between different IDF,
which is traditionally identified with entanglement, or by b)
quantumness of  states of systems with unit number of IDF. The
definition of generalized entanglement assumes that nonminimal
quantumness of noncoherent states of systems with one IDF  is
equivalent to the generalized entanglement. In either the case
$a)$ or $b)$ some Bell inequality
 for a convenient choice of observables is violated by a superposition of $g$-coherent states, that is by generalized entangled states.

Previous discussion in the case when the dynamical group satisfies $G=G^A\otimes G^B$ is generalized by definition to the general case of the systems
 with dynamical groups $G$ such that the decomposition $G=G^A\times G^B$ does not exist.
 Although the system with such $g$ dynamical algebra might have
 more than one IDF it can not be considered as a union of systems with smaller number of IDF.
Nevertheless, the $g$-coherent states are defined and constructed as in
 the general case. The quantumness $\Delta(\psi)$ is minimal for such coherent states and larger than minimal otherwise.
 States  which are not $g$-coherent have the quantumness larger than minimal and are by definition generalized entangled or $g$-entangled states.
Quantumness of the state $|\psi>$: $\Delta(\psi)$, normalized so that it is zero for the $g$-coherent states can
 be used as a measure of $g$-entanglement. It was shown in \cite{Viola3} that it is related to the Mayer-Wallach
 $Q$-measure of multi-partite entanglement in
 the standard case.

Identification of $g$-coherent states with $g$-disentangled states in the case when the dynamical group does not satisfy $G=G^A\otimes G^B$
 should not be
 questionable. Whether a state should be considered as $g$-entangled whenever it is not $g$-coherent is a deep question with no general agrement
  as to
 the answer \cite{Klyasko}. Following \cite{Viola1,Viola2,Viola3} we adopt the identification of $g$-entanglement with $g$-noncoherence.
  This reduces to the standard definition
 in the
 case when $A$ and $B$ have no entangled states and $G=G^A\otimes G^B$.

 If this definition of $g$-entanglement is
 adopted than quantum integrability and $g$-entanglement are clearly related as is explained in the next subsection.

\subsection{ Entanglement generator and integrability}

The system  $(H,g,{\cal H})$ is called an entanglement generator if it does not have $g$ dynamical symmetry. The name is justified because
 an entanglement generator evolves from g-coherent into g-noncoherent states i.e. from g-disentangled into g-entangled states.
Such systems produce entanglement by internal dynamics. On the
other hand, if $(H,g,{\cal H})$ is quantum integrable than the set
of $\{{\rm g-coherent}\}\equiv \{{\rm g-disentangled}\}$ is
dynamically and $G$-invariant.   In this case the Hamiltonian is
not entanglement generator and such a system can be in an
entangled state  as a result of interaction with external systems.
In other words dynamical separability can be identified with
disentanglement.
 This properties provide an understanding of the relation between dynamical integrability and entanglement in quantum mechanics
  and is the main conclusion of our discussion.

\section{Examples}

\subsection{ von-Neumann case: $u(N)$ dynamical algebra}

The quantum system is described by $N$ dimensional Hilbert space $H^N$ and the dynamical algebra $u(N)$, which means that every hermitian operator
 on  $H^N$ has physical interpretation as a measurable quantity. Due to the normalization and global phase invariance the state space of the system
 is $CP^{N-1}$ which is topologically like $S^{2N-1}/S^1$, and represents a $2(N-1)$ manifold with Riemanien and symplectic structure. Geometrically,
 it should be natural to associate $N-1$ IDF with this system. The same number of IDF follows from $u(N)$ dynamical algebra. The Hilbert space
 is the fully symmetric irrep space of $u(N)$ with the highest weight: $\Lambda=(1,0,\dots 0)$. The basis can be labeled by the following chain
 of subalgebras: $u(N)\supset u(N-1)\dots\supset u(1)$ with the corresponding  Casimir operators $C_i^{u(k)},\> i=1,2\dots k, \>k=1,2\dots N$ determine
 the irrep $\Lambda=(1,0,\dots 0)$. The
$N-1$ non-fully degenerate operators are $C_i^{u(k)},\> i=1,2\dots
k, \>k=1,2\dots N-1$ and label the basis $|i>=|0,0,\dots i,\dots
0>,i=0,1,2,\dots N-1$.
Explicitly:$C_k^{u(k)}|i>=\Theta(k-(N-i))|i>$, and $\Theta(i)$ is
the Heaviside function on $i=1,2\dots N-1$. Thus there is $N-1$
IDF, the same as the number of kinematical
 DF.

Any Hamiltonian can be diagonalized by an $U(N)$ transformation and thus expressed as a combination of the Casimir operators. Thus any
 system with $u(N)$ dynamical algebra is quantum integrable.  The classical model for any quantum system with $u(N)$ dynamical
 algebra is also completely integrable when considered as a classical Hamiltonian system.

Elementary excitation operators are given by:
$E_{i0}|\psi_0>=|i>,i=1,2,\dots N-1$ where $|\psi_0>$ is the
lowest weight vector of the $\Lambda=(1,0,\dots 0)$
representation, and $U(N)$ coherent states are obtained as
$|\alpha>=exp(\sum \alpha_i E_{i1}-h.c)|0>$. Coherent states are
parameterized by the coset space $U(N)/U(N-1)\otimes U(1)$ which
is isomorphic to $CP^{N-1}$. We see that all
 states are $U(N)$ coherent states. Thus, all states are equally and minimally quantum.
  The $N-1$ degrees of freedom are disentangled in any
 state.

 It should be noticed that since any state is $u(N)$ coherent
 state the dynamics of the quantum system on $CP^{N-1}$ and its classical
 model with the Hamiltonian function $<{\cal H}>$ on the phase
 space $U(N)/U(N-1)\otimes U(1)$ are identical (and integrable)
 for any Hamiltonian (please see for example \cite{AnnPhys}.

A special case of the systems with $su(N)$ dynamical group is a
qubit with $su(2)$ dynamical algebra and the Hilbert space with
two complex dimensions. The number
 of degrees of freedom of the qubit is one, and all states, like in the general $u(N)$ case, are coherent and equally and minimally quantum.

Systems with $su(2)$ dynamical algebra but with the Hilbert space with $dim> 2$ are treated next.

\subsection{ Entanglement and quantum nonintegrability in a system
with one IDF: $su(2)$ dynamical algebra with $dim H=2j+1>2$}

The two Casimir operators in the subalgebra chain: $su(2)\supset
u(1)$ are $J^2$ and $J_0$. The system has only one IDF,  given by
the only one non-fully
 degenerate operator $J_0$.
Hamiltonian which is a linear expression of the $SU(2)$ generators
is quantum integrable according to the definition (with the proper
choice of the quantization axes). A system with a Hamiltonian that
is a nonlinear expression of the generators is quantum
nonintegrable, and as we shall see
 generates $g$-entanglement.

The $SU(2)$ coherent states are given by: $|\alpha>=exp( (\alpha J_{+}-h.c)|0>$ where $|0>$ is the unique lowest weight
 vector in the representation $ H^{2j+1}$ and $J_{+}$ is the corresponding raising operator. States which are not coherent are more
 quantum in the sense
 that they have larger $\Delta$ than the coherent states. According to the accepted definition such states are $g$-entangled.

 If the Hamiltonian is a linear expression in terms of the $su(2)$ generators, i.e. an element of the $su(2)$ algebra then the set of coherent
 states is dynamically invariant.
On the other hand, when the Hamiltonian is a nonlinear expression
of the $su(2)$ generators the states with different levels of
quantumness are not dynamically isolated as is illustrated in fig.
1. Provided the accepted definitions of quantum nonintegrability
and $g$-entanglement we see that
 nonintegrable Hamiltonians generate $g$-entanglement in the systems with one
 IDF. The data presented in fig. 1. are generated using the
 Hamiltonian
 \begin{equation}
 {\cal H}=\omega_z J_z-2\omega_x J_x+\mu J_z^2,
 \end{equation}
  which
 is integrable when $\mu=0$ and nonintegrable when $\mu\neq 0$.

\subsection{ Coupled spins: $su^1(2)\oplus su^2(2)$ dynamical
algebra}

Consider a pair of spins with the Hilbert space $H=H^{2j_1+1}\otimes H^{2j_2+1}$ and the Hamiltonian
\begin{equation}
 H=(1-\mu)(J_z^1+J_z^2)+\mu J_x^1J_x^2,
\end{equation}
where $\mu=\neq 1$. The case $\mu=1$ is treated separately.

 The dynamical group of the system is $SU^1(2)\otimes SU^2(2)$. The
  subgroup chain $\alpha: \> SU^1(2)\otimes SU^2(2)\supset SO^1(2)\otimes SO^2(2)$ gives two IDF  and the Casimir operators of the subgroups
$J_z^1$ and $J_z^2$ are the observables corresponding to the two
IDF.

The system is quantum  integrable  when $\mu=0$ because of the $SU^1(2)\otimes SU^2(2)$ dynamical symmetry.
 If $\mu\neq 0$ and $\mu\neq 1$ the system is quantum nonintegrable. As was already pointed out, the orbits in the Hilbert space
  of the quantum integrable
 and nonintegrable cases belong in the same class from the
 point of view of the qualitative theory of dynamical systems, i.e. they are regular orbits.

Because of the definition of the dynamical group as  $SU^1(2)\otimes SU^2(2)$ the system is considered as composed of two spins.
$SU^1(2)\otimes SU^2(2)$.
 Coherent states  are products of the coherent states of each of the spins and are thus disentangled. If $j_1>1/2$ or $j_2>1/2$ then
  there are product states that are products of noncoherent states of the component spins. They are noncoherent but product states. These
  posses larger then minimal quantumness and  we are considering such states as  $g$-entangled.
If the system is quantum integrable due to the symmetry $\alpha$ the $SU^1(2)\otimes SU^2(2)$ coherent i.e. disentangled states are dynamically
 invariant. Likewise, the set of noncoherent states i.e. $g$-entangled states is also dynamically invariant.
These sets are also $SU^1(2)\otimes SU^2(2)$ invariant. If $\mu\neq 0$
 the system is not $SO^1(2)\otimes SO^2(2)$ quantum integrable and the sets of coherent i.e  $g$ disentangled and nonhoherent i.e.
  $g$-entangled states
 are not dynamically invariant. The system generates entanglement between the  $SO^1(2)\otimes SO^2(2)$
 dynamical degrees of freedom (please see fig. 2 ).

Consider now the special case $\mu=1$. It is more natural to consider the group $SU^{1+2}(2)$ as the dynamical group of the system, with
 the subgroup
 chain $SU^{1+2}(2)\supset SO^{1+2}(2)$, with the $x$-axis as the $SO(2)$ axis. The Hilbert space is  a sum of $j=0,j=1$ $SU^{1+2}(2)$
 irrep spaces.
 The system should be considered as one degree of freedom and is quantum integrable. The level of quantumness
  is preserved by the evolution with the Hamiltonian for $\mu=1$.
  Special to this case is the fact that the disentangled
  $|1/2,-1/2>\otimes |1/2,-1/2>=|1,-1>$ and the maximally
  entangled $(|1/2,-1/2>\otimes |1/2,1/2>+|1/2,1/2>\otimes
  |1/2,-1/2>)/\sqrt 2=|1,0>$ states are the same states in $SU^1(2)\otimes
  SU^2(2)$ or $SU^{1+2}(2)\supset SO^{1+2}(2)$ choices of the IDF,
  despite the fact that the number of IDF is two or one,
  respectively. In general, the set of $g$-disentangled and
  $g$-entangled states with respect to different IDF are
  different.

\subsection{ A simple system with entanglement: $su(3)$ dynamical
algebra}

The example of $su(3)$ dynamical algebra is used to illustrate the
systems with more than one IDF which nevertheless can not be
considered as composed of component systems with fewer number of
IDF because the Hilbert space of states does not have the
corresponding tensor product structure. The example will also
illustrate another important fact, namely the fact that the number
of IDF might depend on the particular irrep that is carried by
 the system's Hilbert space.

The $su(2)$ Lie algebra has rank $2$ and dimension $8$. The basic commutation relations between the generators $E_{i,j}, \>i,j=1,2,3$ which are
 not independent are: $[E_{ij},E_{kl}]=\delta_{jk}E_{il}-\delta_{il}E_{kj}$ and can be realized
 in terms of bosonic creation and annihilation operators of three modes as follows: $E_{i,j}=a_i^{\dag},a_j,\>i,j=1,2,3$.
 The eight independent  hermitian generators are given by:
 $X_1=(a_1^{\dag}a_1 -a_2^{\dag}a_2);\>X_2=(a_1^{\dag}a_1
 -a_2^{\dag}a_2-2a_3^{\dag}a_3);\>Y_k=i(a_k^{\dag}a_j-a_j^{\dag}a_k);\>Z_k=(a_k^{\dag}a_j-a_j^{\dag}a_k),\>
 k=1,2,3,\> j=k+1 {\rm mod} 3$. These will be  used in the formula
 (1) for the level of $g$-entanglement.

In order to determine the number of IDF we need to find the number
of nonfully degenerate operators in any  particular chain of
subalgebras. We shall use the subalgebra chain: $su(3)\supset
su(2)\oplus u(1)\supset u(1)$ with five Casimir operators usually
denoted by $C_2,C_3,Y,T^2,T_z$. $C_2$ and $C_3$ are the Casimir
operators of the su(3) itself, $T^2$ and $T_z$ are the Casimir
operators of $su(2)$ and u(1) and $Y$ corresponds to $u(1)$. In
the most famous application, that is in the $SU(3)$ quark model,
operators $Y$ and $T^2$,$T_z$ represent hypercharge, isospin and
its $z$ component. Thus, in general there are three nonfully
degenerate operator and consequently a system with $su(3)$ algebra
has three
 IDF. However, the system is also characterized by its Hilbert space i.e. by a particular irrep and for some irrep all three DF might not
  be independent.

All irreps of the $su(3)$ algebra can be labeled by their highest
weight: $\Lambda=\lambda_1f_1+\lambda_2 f_2$ where $f_1$ and $f_2$
are the highest
 weights of the two fundamental representations: $(1,0)$ and $(0,1)$.  The fully symmetric representations correspond to
 $\lambda_1=0$ or $\lambda_2=0$.
In the fully symmetric representation the operators $T^2$ and $Y$
are not independent and thus in this case the number of IDF is
just 2. A particular
 Hamiltonian is quantum integrable if it is expressed in terms of $T^2$ and $T_z$  or $Y,T^2$ and $T_z$ in the two or
  three degrees of freedom cases.
Hamiltonians used in the $SU(3)$ quark model are integrable by
construction.

The coherent states of the $SU(3)$ dynamical group are obtained as in the general case using the highest weight vector as the reference state $|\psi_0>$.
 In the general case the coherent states are parameterized by the six dimensional manifold: $SU(3)/U(1)\otimes U(1)$ and in the case of
 the fully symmetric irrep with two IDF by the four dimensional $SU(3)/U(2)$. As usual the coherent states are of the form
$|\Lambda,\alpha>= D(\alpha)|\psi_0>$. According to the adopted definition the coherent states are disentangled and quantum integrable Hamiltonian
 systems, like those of the quark model, can not generate entanglement. Quantum $su(3)$ nonintegrable Hamiltonians and the generation of
$su(3)$ generalized entanglement by the system's dynamics is illustrated in the following example.

Consider the system of $N$ particles with three possible
$N_d$-degenerate energy levels. The following Hamiltonian for such
a system is known
 as the Lipkin model:
\begin{equation}
H=\sum_{i=1}^3\omega_i E_{ii}-\mu\sum_{i\neq j}^3E_{i,j}^2
\end{equation}
where $E_{ij}$ satisfy $su(3)$ commutation relations. Quantum
integrability and nonintegrability of this system was studied in
\cite{Cheng2}, and the
 chaotic dynamics of the classical limit was analyzed in \cite{Mer}.
 When $N\leq N_d$ the Hilbert space of the system is the carrier space of the fully-symmetric
 irrep and the system has two IDF. If $\mu=0$ there is the dynamical symmetry corresponding to
  $su(3)\supset su(2)\oplus u(1)\supset u(1)\times u(1)$,
 the system is quantum integrable and does not generate $g$-entanglement.
 For $\mu\neq 0\neq \omega_i$ the system is quantum nonintegrable and does
 generate $g$-entanglement (please see fig. 3).

If $N>N_d$ the constraints imposed on the Hilbert space by the Pauli principle become important.

 \subsection { N level system of fermions with $U(N)$ dynamical algebra}

There are several dynamical algebras that can be constructed from
the creation and annihilation operators of a system of identical
fermions or bosons. The dynamical algebra of a system of $k$
identical particles distributed on $N$ levels can be chosen to be
$u(N)$, with different Hilbert spaces of states
  for fermions and for bosons. In the fermion case the Hilbert space is the carrier space of the fully antisymmetric representation of $u(N)$ denoted
 by $\Lambda=(\lambda_1,\lambda_2\dots \lambda_N)=(1,1,\dots 1,0,0\dots 0)=(1^N,0^{N-k})$. The basis states of this $d=N!/k!(N-k)!$ dimensional Hilbert space are
of the form $|n_1,n_2\dots n_N>$ with $\sum_{i=1}^N n_i=k$, and
the number of degrees of freedom is $N(N-k)$.

The extremal state of the fully antisymmetric representation to be
used for the construction of the coherent states is the state
labeled $|\psi_0>=|1^N,0^{N-k}>$, which is actually the ground
state of the unperturbed many-fermion system. With the standard
 notation for the fermions creation and annihilation operators: $a_i^{\dag}, a_j,\>i,j=1,2\dots N$ the operators
 $a_i^{\dag}a_j,\> i\leq k, k+1\leq j\leq N$  generate the subgroup $U(k)\otimes U(N-k)$ which is the stability subgroup of the reference state
 $|\psi_0>$.
The coherent states are of the form: $|\Lambda,\alpha>=\sum
exp(\alpha_i a_i^{\dag}a_j+h.c )|\psi_0>$,
 where the sum extends over $k+1\leq i\leq N,\> 1\leq j\leq k$.
By the adopted definition, these are $g$-disentangled states of the fermion system, and the noncoherent states are $g$-entangled.

Let us consider a many fermion system with the following Hamiltonian: $H=\sum _i^N \omega_i a_i^{\dag}a_i+\mu V_{int}$. When $\mu=0$, corresponding
 to the noninteracting fermions the system is
 quantum integrable since it has the dynamical symmetry, i.e. the Hamiltonian is expressed in terms of the Casimir operators
 of the subgroup chain: $U(N)\supset\dots\supset U(1)\otimes U(1)\otimes\dots\otimes U(1)\supset SO(3)\supset SO(2)$. The coherent states are an invariant set for
 the evolution generated by this Hamiltonian, and such evolution does not generate $g$-entanglement. Of course,
 an analysis  of the interacting systems could be to complicated but in general they are quantum nonintegrable and generate the $g$-entanglement.

 \section{Summary}

We have used the dynamical algebra definition of independent
degrees of freedom in order to establish a general relation
between quantum integrability  or nonintegrability and the
dynamics of the generalized entanglement (g-entanglement).
Generally applicable definition of degrees of freedom of a quantum
system requires specification of the system's dynamical algebra,
which physically corresponds to the set of measurable observables
of the system. Quantum integrability is identified with dynamical
symmetry with respect to the algebra used to define the degrees of
freedom. Minimal level of total quantum fluctuations is a property
characteristic of the dynamical algebra generalized coherent
states. States with non-minimal quantum fluctuations are here
identified (following \cite{Viola1,Viola2,Viola3}) with the
g-entangled states. With this identification, both sets of
g-disentangled and g-entangled states are dynamically invariant
for the quantum integrable systems. On the other hand, an orbit of
the quantum nonintegrable system goes through states with zero and
nonzero g-entanglement. Quantum nonintegrable systems generate g-
entanglement by the internal dynamics, while quantum integrable
systems can be in a g-entangled state only due to interactions
with external systems. The relation between dynamical symmetry and
g-entanglement is manifested also in the relation between chaotic
dynamics of the quantum system's classical model and dynamical
generation of g-entanglement in the quantum system. Several
examples of the relation between g-entanglement and quantum
nonintegrability have been discussed.

  \vskip 0.5cm

 {\bf Acknowledgements} This work is
partly supported by the Serbian Ministry of Science contract No.
141003. I  would also like to acknowledge  the support of Abdus
Salam ICTP.

\newpage

\begin{figure}
 \includegraphics[height=0.5\textheight,width=0.9\textwidth]{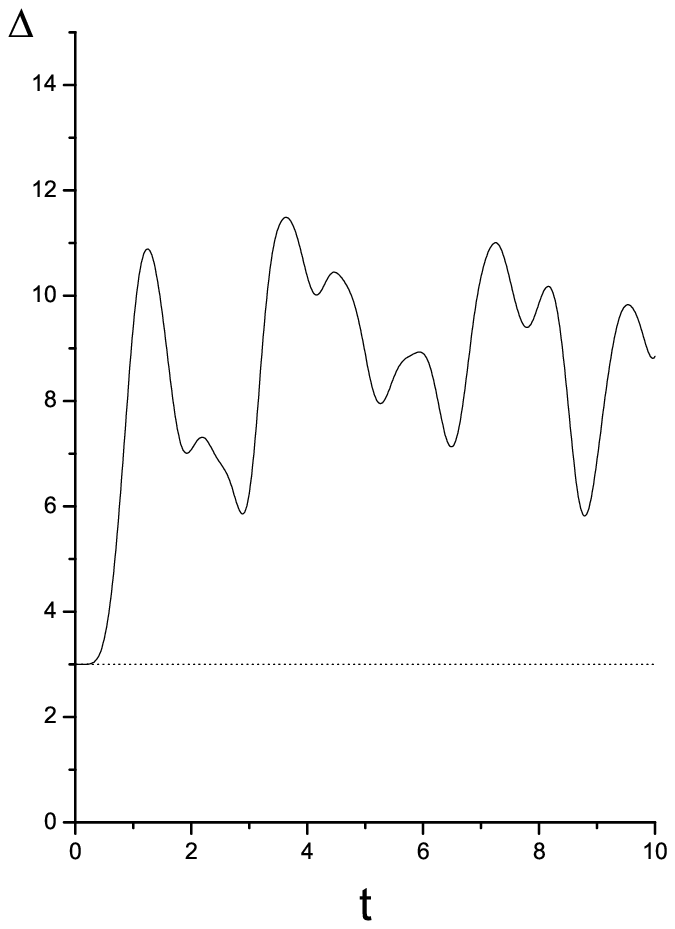}
\caption{ Illustrates dynamics of the total quantum fluctuation
$\Delta(t)$ with the Hamiltonian (3), starting from an $SU(2)$
coherent states in $j=3$ irrep. Full line corresponds to the
quantum nonintegrable $\mu=1,\omega_x=\omega_z=1, $ and dotted to
the integrable cases $\mu=0,\omega_x=0,\omega_z=1$ and
$\mu=0,\omega_x=1,\omega_z=1$.}
\end{figure}
\newpage

\begin{figure}
 \includegraphics[height=0.5\textheight,width=0.9\textwidth]{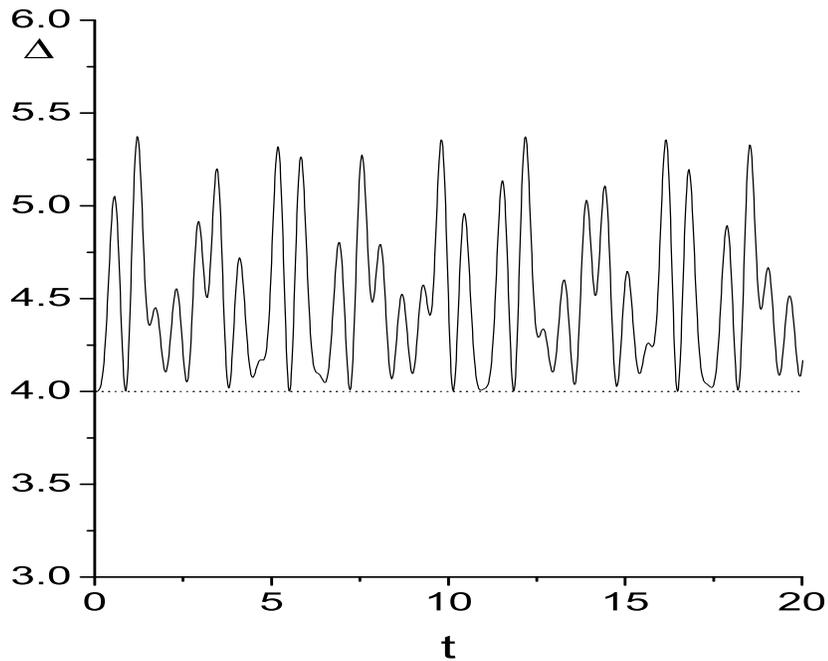}
\caption{  Illustrates dynamics of the total quantum fluctuation
$\Delta(t)$ with the Hamiltonian (4), starting from an
$SU(2)\otimes SU(2)$ coherent states in $1/2\otimes 1/2$ irrep.
Full line corresponds to the quantum nonintegrable $\mu\neq 0,1$
and dotted to the integrable case $\mu=0$. In the nonintegrable
case the line $\Delta(t)$ comes very close but remains larger than
the initial value.}
\end{figure}
\newpage
\begin{figure}
 \includegraphics[height=0.5\textheight,width=0.9\textwidth]{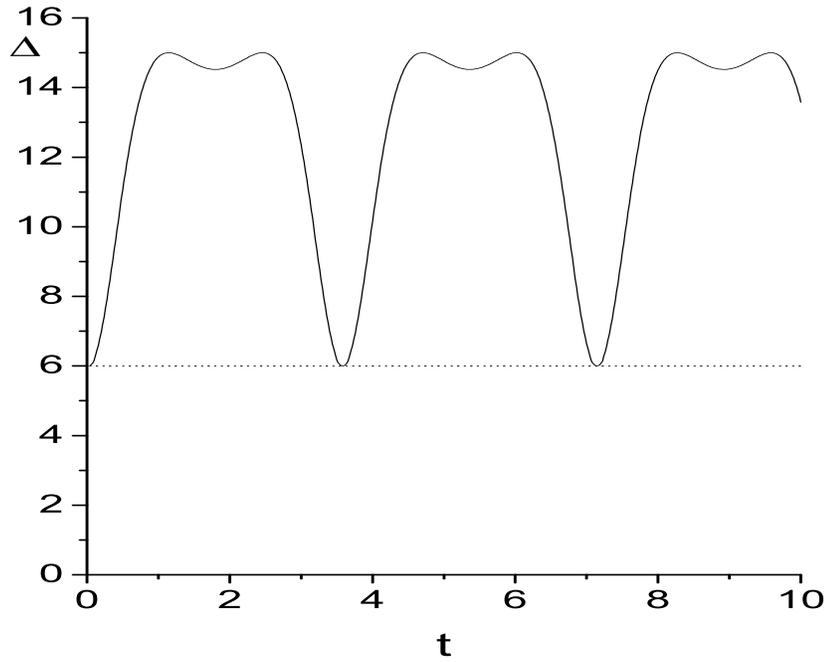}
\caption{ Illustrates dynamics of the total quantum fluctuation
$\Delta(t)$ with the Hamiltonian (5), starting from an $SU(3)$
coherent states in the completely symmetric irrep. Full line
corresponds to the quantum nonintegrable $\mu=1/6,\omega_i=1$ and
dotted to the integrable case $\mu=0,\omega_i=1$. In the
nonintegrable case the line $\Delta(t)$ comes very close but
remains larger than the initial value.}
\end{figure}

\end{document}